\documentclass[aps,notitlepage,amsmath,amssymb,superscriptaddress,onecolumn]{revtex4-1}

\usepackage{graphicx} \usepackage{color} \usepackage{verbatim} 
\usepackage[colorlinks, citecolor={blue}, urlcolor={blue}, linkcolor={blue}]{hyperref} 
\usepackage{comment} 

\definecolor{darkgreen}{rgb}{0,0.5,0}
\definecolor{purple}{rgb}{0.6,0,0.5}
\definecolor{orange}{rgb}{1,0.5,0}
\definecolor{darkred}{rgb}{.7,0,0}
\definecolor{darkblue}{rgb}{0,0,.3}
\definecolor{grey}{rgb}{.6,.6,.6}
\definecolor{dimgreen}{rgb}{0.2,0.6,0.1}

\usepackage{scalefnt}
\newcommand{\aheading}[1]{\medskip\noindent{\scalefont{1.05}{\textbf{#1}}}}
\newcommand{\bheading}[1]{\noindent\textbf{#1.}}

\newcommand{\ks}[1]{{\color{black}{#1}}} 
\newcommand{\gk}[1]{{\color{black}{#1}}} 
\newcommand{\jvd}[1]{{\color{black}{#1}}} 
\newcommand{\JvD}[1]{{\color{black}{#1}}} 
\newcommand{\JVD}[1]{{\color{black}{#1}}} 
\newcommand{\jVd}[1]{{\color{black}{#1}}} 
\newcommand{\interact}{{\rm int}}

\newcommand{\free}{{\rm free}}
\newcommand{\spin}{{\rm spin}}
\newcommand{\orb}{{\rm orb}}

\newcommand{\pdag}{{\phantom{\dagger}}}

\newcommand{\chispin}{\chi_{\rm{spin}}}
\newcommand{\chiorb}{\chi_{\rm{orb}}}
\newcommand{\chicharge}{\chi_{\rm{charge}}}

\renewcommand{\section}{\flushleft\thesection}

  \usepackage[normalem]{ulem}
  \usepackage{xspace}

  \def\Eatomic{\ensuremath{E_{\mathrm{atomic}}}\xspace}
  \def\TspinO{\ensuremath{T_{\mathrm{spin}}^{\mathrm{onset}}}\xspace}
  \def\TspinC{\ensuremath{T_{\mathrm{spin}}^{\mathrm{cmp}}}\xspace}
  \def\TKspin{\ensuremath{T^{\mathrm{spin}}_{\mathrm{K}}}\xspace}
 \def\TFL{\ensuremath{T_{\mathrm{FL}}}\xspace}

  \def\TorbO{\ensuremath{T_{\orb}^{\mathrm{onset}}}\xspace}
  \def\TorbC{\ensuremath{T_{\orb}^{\mathrm{cmp}}}\xspace}
  \def\TKorb{\ensuremath{T^{\mathrm{\orb}}_{\mathrm{K}}}\xspace}

  \def\TM{\ensuremath{T_{\mathrm{M}}}\xspace}

\newcommand{\vo}{V$_2$O$_3$}

\newcommand{\sroone}{Sr$_2$RuO$_4$}

\def\K{\mathrm{K}}
\def\eV{\mathrm{eV}}

\begin{document}

\title{Signatures of Mottness and Hundness in archetypal correlated metals}

\author{Xiaoyu Deng}
  \thanks{These authors contributed equally to this work. Correspondence: \href{mailto:xiaoyu.deng@gmail.com}{xiaoyu.deng@gmail.com}.}
   \affiliation{Department of Physics and Astronomy, Rutgers University,
  Piscataway, New Jersey 08854, USA}


%

\author{Katharina M. Stadler}
  \thanks{These authors contributed equally to this work. Correspondence: \href{mailto:xiaoyu.deng@gmail.com}{xiaoyu.deng@gmail.com}.}
\affiliation{Physics Department, Arnold Sommerfeld Center for Theoretical Physics and Center for NanoScience,
Ludwig-Maximilians-Universitat M\"unchen, 80333 M\"unchen, Germany}

\author{Kristjan Haule}
\affiliation{Department of Physics and Astronomy, Rutgers University,
  Piscataway, New Jersey 08854, USA}

\author{Andreas  Weichselbaum}
\affiliation{Condensed Matter Physics and Materials Science Department,
  Brookhaven National Laboratory, Upton, New York 11973, USA}
\affiliation{Physics Department, Arnold Sommerfeld Center for Theoretical Physics and Center for NanoScience,
Ludwig-Maximilians-Universitat M\"unchen, 80333 M\"unchen, Germany}

\author{Jan  von  Delft}
\affiliation{Physics Department, Arnold Sommerfeld Center for Theoretical Physics and Center for NanoScience,
Ludwig-Maximilians-Universitat M\"unchen, 80333 M\"unchen, Germany}

\author{Gabriel Kotliar}
\affiliation{Department of Physics and Astronomy, Rutgers University,
  Piscataway, New Jersey 08854, USA}
\affiliation{Condensed Matter Physics and Materials Science Department,
  Brookhaven National Laboratory, Upton, New York 11973, USA}

\begin{abstract}
  Physical properties of multi-orbital materials depend not only on
  the strength of the effective interactions among the valence
  electrons but also on their type. Strong correlations are caused by
  either Mott physics that captures the Coulomb repulsion among
  charges, or Hund physics that aligns the spins in different
  orbitals.  We identify four energy scales marking the onset and
    the completion of screening in orbital and spin channels.  The
    differences in these scales, which are manifest in the temperature
    dependence of the local spectrum and of the charge, spin and
    orbital susceptibilities, provide clear signatures 
    distinguishing Mott and Hund physics. We illustrate these concepts
    with realistic studies of two archetypal strongly correlated
    materials, and corroborate the generality of our conclusions with a
    model Hamiltonian study.
\end{abstract}

\date{\today}
  
\maketitle

The excitation spectra and transport properties of transition metal
oxides at high energy and/or high temperature are well described in
terms of dressed atomic excitations with their characteristic
multiplet structure. At very low energy scales, \JVD{by contrast, metallic
  systems are well described} in terms of strongly renormalized Landau
quasiparticles forming dispersive bands.  Describing the evolution of
the excitation spectrum as a function of energy scale is a fundamental
problem in the theory of strongly correlated materials. Starting with
the Fermi liquid quasiparticles at the lowest energy scales, and
raising the temperature, one can view this evolution as their gradual
undressing. Conversely, starting from the high energy end, one can
understand the evolution of the excitation spectrum as the 
\JVD{screening} of the orbital and spin excitations of atomic states,
which gradually bind to give rise to quasiparticles. 
\JVD{Here we consider the temperature dependence of 
this screening process for correlated multi-orbital systems with
strong on-site atomic-like interactions, involving}
\JVD{both a Coulomb repulsion $U$ and Hund's coupling $J$. The former
  differentiates between different charge configurations without
  preference for a given spin or orbital configuration, whereas the
  latter favors the highest spin state.}  

It is well known that strong
correlation effects can arise due to proximate Mott insulating states
in which strong on-site Coulomb repulsion slows down charge
fluctuations or even blocks the charge motion and localizes the
electrons \cite{marel1988,imada1998}.  However, many materials far
away from the Mott insulating state, notably the $3d$ iron-based
superconductors \cite{haule2009c,yin2011a} and ruthenates
\cite{mravlje2011,werner2008}, display strong correlation effects as a
result of strong Hund coupling rather than the Hubbard $U$.  These
so-called ``Hund metals"
\cite{haule2009c,yin2011a,mravlje2011,werner2008,medici2011,georges2013,
  yin2012,khajetoorians2015,aron2015, stadler2015, stadler2018,
  mravlje2016} were proposed to be a new type of strongly correlated
electron system, characterized by spin-orbital separation
\cite{yin2012,stadler2015, stadler2018}.  

\gk{The existence of
different origins of correlations, \jVd{Coulomb $U$ or Hund $J$}, 
poses an important questions: what are the defining signatures 
\jVd{distinguishing Mott and Hund metals?}}  \jvd{The goal of
  this work is to answer this question, by pointing out that Hund and
  Mott metals differ strikingly in the temperature dependencies of
  their local correlated spectra and the local susceptibilities
  describing the charge, spin and orbital degrees of freedom. These
  differences reflect two \JVD{distinct} screening routes for how
  quasiparticles emerge from the atomic degrees of freedom, with
  spin-orbital separation involved for Hund metals, but not for Mott
  metals.}

\JvD{We provide evidence of the two distinct screening routes by
  investigating (i) two realistic materials and (ii) a model
  Hamiltonian.  For (i) we consider two archetypical materials with
  non-degenerate orbitals, the Mott system \vo\
  \cite{mcwhan1969,mcwhan1973a,mcwhan1973} and the Hund metal \sroone\
  \cite{maeno1994}. We compute their properties using density
  functional theory plus dynamical mean-field theory (DFT+DMFT)
  \cite{georges1996,kotliar2006,held2007}, which has been successfully
  used to describe the available experimental measurements for \vo\
  \cite{held2001,laad2006,poteryaev2007,hansmann2013,
    grieger2014,grieger2015,deng2014} and \sroone\
  \cite{mravlje2011,dang2014,dang2015,deng2016}.  For (ii) we study a
  3-band Hubbard-Hund model (3HHM) with three degenerate bands hosting
  two electrons. This 3HHM is the simplest model capable of capturing
  both Hund and Mott physics and the crossover between them \jVd{as
    function of increasing $U$} \cite{stadler2015, stadler2018}.
  Whereas Ref.~\cite{stadler2018} focused on $T=0$, here we focus on
  temperature dependence.
  We study the 3HHM using DMFT and the numerical renormalization group
  (DMFT+NRG).  We accurately determine the location of the Mott
  transition at zero temperature and show \JVD{that provided that $J$
    is sizeable, the temperature dependence of physical properties for large $U$ near the Mott
    transition line qualitatively resembles that of \vo, while for
    small $U$ far from the transition it resembles that of
    \sroone. Therefore, our 3HHM results elucidate the physical origin of
    the differences between these materials.  Indeed, we argue that 
our findings are} applicable to general multi-orbital materials and
  characteristic of the general phenomenology of Mott and Hund
  physics, independent of material-dependent details, such as the
  initial band structure.}

\begin{figure}
\includegraphics[width=0.7\columnwidth]{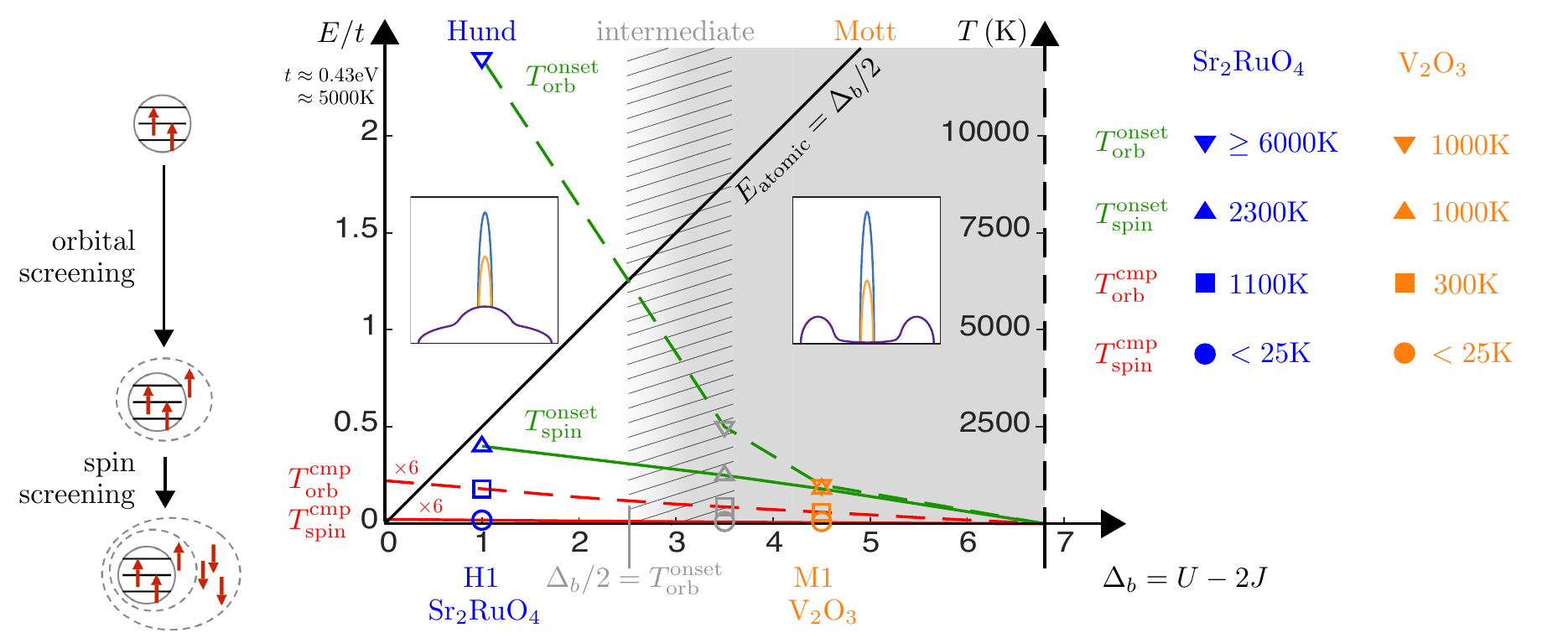}
\caption{\jvd{Schematic sketch of the behavior of four characteristic
    temperature scales, $\TorbO$ (green dashed), $\TspinO$ (green
    solid), $\TorbC$ (red dashed), $\TspinC$ (red solid), marking the
    onset and the completion of screening of orbital and spin degrees
    of freedom, respectively, as functions of the bare gap,
    $\Delta_b = U - 2J$, between the upper and lower Hubbard side
    band.  Open symbols on the left give the values of these scales as
    obtained from DMFT+NRG calculations for our 3-band Hubbard-Hund
    model, with $J=1$ and $U=3$ (green), $U=5.5$ (grey) and $U=6.5$
    (yellow). On the right, closed symbols give corresponding values
    obtained from DFT+DMFT calculations for the materials \sroone\
    (green) and \vo\ (yellow). \JVD{Left: cartoon of two-stage
      screening for three local levels with $J \neq 0$, containing two
      electrons with total spin $S=1$: with decreasing temperature,
      first orbital screening of the hole occurs, whereby a
      delocalized spin $1/2$ combines with the local spin $1$ to
      yield an orbital singlet with spin $3/2$; then spin screening
      occurs, yielding an orbital and spin singlet \cite{stadler2018}.  The energy scales
      characterizing the two screening stages lie far apart for Hund
      systems, but close together for Mott systems.} Insets: cartoons
      of the local density of states, $A(\omega)$, for a Hund system
      (left) and a Mott system close to the Mott transition (right),
      summarizing the essential differences in the evolution of the
      quasiparticle peak with decreasing temperature (purple to yellow
      to blue).  }}
\label{sketch}
\end{figure}

\JvD{\aheading{Overview of results} 


\noindent 
We start with  an overview of our most important observations.
We first identify four temperature scales, characterizing the onset
and completion of screening of the orbital and spin degrees of freedom
as the temperature is lowered.  The scales for the onset of screening,
\TorbO and \TspinO, are defined as the temperatures at which the
static local orbital and spin susceptibilities, $\chiorb$ and
$\chispin$, first show deviations from the Curie behavior,
$\chi \propto 1/T$, shown by free local moments.  The scales for the
completion of screening, \TorbC and \TspinC, mark the transition of
these susceptibilities to Pauli behavior, saturating to constants at
very low temperatures.}  \jvd{For orientation, Fig.~\ref{sketch}
summarizes the behavior of these scales with increasing Coulomb
interaction at fixed Hund's coupling, as will be elaborated throughout
the text below.  \jVd{The most striking observation is that increasing $U$
pushes the onset scales $\TorbO$ and $\TspinO$ closer together until
they essentially coincide. As a consequence, the Hund regime (small
$U$) and the Mott regime (large $U$, close to the Mott transition),
though adiabatically connected via a crossover regime,
show dramatic differences for the temperature dependence of
physical quantities (discussed below).} The 
trends shown in the figure were extracted from our analysis of the 3HHM, 
but they match those found for \vo\ and \sroone\ (see legend on the right), 
and we expect them to be generic for multi-orbital Mott and/or Hund systems.}

\JvD{\bheading{Mott systems} 
\jvd{Coulomb interactions are strong in
    Mott systems, so that there is a large separation between the
    upper and lower Hubbard bands associated with atomic excitations,
  } \JVD{resulting in a significant gap or pseudogap. It is well
    established that, in the context of a metal-insulator transition,
    a metallic state is induced by the formation of a quasiparticle
    resonance from either the center of the gap or the edges of the
    Hubbard bands.  Vice versa, starting from a correlated metal of
    Mott-type, a gap (or pseudogap) between incoherent spectra is
    restored when the coherence resonance is destroyed.  The \jVd{transition
   from insulator to metal} can
    be induced by doping or tuning the ratio of the interaction versus
    the bandwidth \cite{imada1998}, but it also occurs when the
    coherence resonance \jVd{emerges by decreasing temperature}.
    For example, in a model study of a doped single-band Mott
    insulator at infinite dimensions, the spectra at high temperature
    exhibit a two-peak structure reminiscent of a Mott gap, \jVd{from
      which a coherence resonance emerges} as ``resilient
    quasiparticles" \jVd{appear} \cite{deng2013}.  

Here we propose
    that the \jVd{presence} 
 of a gap/pseudogap regime in the local density
    of states (LDOS) \jVd{at temperatures so high that
the coherence resonance is destroyed}
is a fingerprint of Mott physics in general situations.
We show that, at high temperatures, the LDOS of a multi-orbital Mott}
  system exhibits two incoherent peaks above and below the Fermi
  energy, reflecting atomic particle- and hole-like excitations, with
  a pseudogap in between, signifying the proximity to charge
  localization.  \JVD{When the temperature is lowered, a clear and sharp
  quasiparticle resonance suddenly emerges near the Fermi energy}
  \jvd{(for a cartoon depiction, see Fig.~\ref{sketch}, right inset),}
  signifying the appearance of mobile charge carriers.
\JVD{This occurs at a well-defined temperature scale, $T_M$, well below
  the lowest atomic excitation energy, $\Eatomic$, involved in the
  incoherent two-peak structure. 
  \JVD{This Mott behavior is also reflected in}  
  the temperature dependence of various local susceptibilities.}  
   With decreasing temperature the static
  local charge susceptibility and local charge fluctuations first
  remain small and rather constant while the local spin and orbital
  susceptibilities exhibit Curie behavior, indicative of unscreened
  local moments. Once the temperature drops below $T_M$, 
  the appearance of mobile carriers causes the local charge
  susceptibility and charge fluctuations to increase, and the spin and
  orbital susceptibilities to deviate from pure Curie behavior,
  reflecting the onset of screening. For Mott systems, this onset thus
  occurs simultaneously for orbital and spin degrees of freedom,
  $\TspinO = \TorbO=T_M$.

  \bheading{Hund systems} The above signatures of Mott physics are in
  stark contrast to the behavior of Hund systems. These typically have
  much smaller values of $U$, and hence $\Eatomic$. Consequently, the
  Hubbard side bands effectively overlap, so that the LDOS features a
  single incoherent peak even at temperatures as high as $\Eatomic$ or
  beyond.  This broad peak evolves into a coherent quasiparticle peak
  as the temperature is lowered \jvd{(for a cartoon depiction, see
    Fig.~\ref{sketch}, left inset)}. \JVD{Due to the absence of a
    pseudogap at large temperatures,} the local charge susceptibility
  is large already at high temperatures and increases
  continuously, but only slightly, with decreasing temperature.
  Strikingly, the local orbital and spin susceptibilities show
  \JvD{deviations from} Curie-like behavior already at much higher
  temperatures than those in Mott systems. Moreover, \JvD{orbital
    screening starts well before spin screening,
    $\TorbO \gg \TspinO$.}  Thus, Hund metals \JvD{exhibit
    spin-orbital separation, featuring} a broad temperature window, from
  $\TorbO$ down to $\TspinO$, involving screened, delocalized orbitals
  coupled to unscreened, localized spins. 
  \JVD{Importantly, $\TorbO$ \JVD{can  be} much larger than $\Eatomic$ in Hund
    metals, which is why no pseudogap appears even up to temperatures
    well above $\Eatomic$ (it would appear only for
    $T \gtrsim \TorbO$, since it requires the breakdown of both spin
    \textit{and} orbital screening).}
\jvd{The fact that
  $\TspinO$ and $\Eatomic$ are both $\ll \TorbO$} is a crucial difference relative to
Mott systems. There \JvD{$\TspinO \simeq \TorbO \ll \Eatomic$,} so
that the breakdown of spin and orbital screening, and the concomitant
emergence of a pseudogap, \textit{is} possible at temperatures well
below $\Eatomic$.

\ks{In principle,} both multi-orbital Mott and Hund materials exhibit
spin-orbital separation in the \textit{completion} of screening:
$\chiorb$ crosses over to Pauli behavior at a larger temperature scale
than $\chispin$, i.e.\ $\TorbC \gg \TspinC$. Since Fermi-liquid
behavior occurs below \TspinC, this scale can be identified with the
Fermi-liquid scale $\TFL\equiv \TspinC$.}  \ks{However, spin-orbital
separation in the completion of screening is much more pronounced for
the Hund material.}

\JvD{\aheading{Two archetypical materials: \vo\ and \sroone}
  
\JvD{\bheading{Established facts}}
We begin our discussion of the two example materials by
  summarizing some of their well-established properties.}  \vo, a
paramagnetic metal at ambient conditions, is proximate to an
isostructural Mott transition (that can be induced by slightly
Cr-doping), and a temperature-driven magnetic transition
\cite{mcwhan1969, mcwhan1973a,mcwhan1973}. It exhibits Fermi-liquid
behavior at low temperature when antiferromagnetism is quenched by
doping or pressure \cite{mcwhan1969,mcwhan1973a,mcwhan1973}.  \sroone,
on the other hand, is a paramagnetic metal far away from a Mott
insulating state \cite{carlo2012a}. As temperature decreases it shows
Fermi-liquid behavior and eventually becomes superconducting at very
low temperature \cite{maeno1994}. Despite the very different distances
to a Mott insulating state, both materials have large specific heat
coefficients in their Fermi-liquid states
\cite{mcwhan1969,mcwhan1973a, mcwhan1973, maeno1994}. In both
materials the observed Fermi-liquid scales are extremely low (around
$25\K$ \cite{mcwhan1969,mcwhan1973a,mcwhan1973,hussey1998}), much
smaller than the bare band energy or interaction parameters (order of
$\eV$). Pronounced quasiparticle peaks are observed in both materials
using photoemission spectroscopy
\cite{yokoya1996,mo2003,rodolakis2009,fujiwara2011}, and large values
of mass renormalization are seen in \sroone\ in various measurements
\cite{bergemann2003,iwasawa2012,veenstra2013}.  Notably, the local
physics on V/Ru sites are similar, with nominally two electrons/holes
in three $t_{2g}$ orbitals.  Due to the crystal field of the
surrounding oxygen, the $t_{2g}$ orbitals of V are split into
$e^{\pi}_{g}$ orbitals with two-fold degeneracy and an energetically
higher-lying $a_{1g}$ orbital, while those of Ru are split into
$xz/yz$ orbitals with two-fold degeneracy and an energetically
lower-lying $xy$ orbital.  Two electrons (holes) in three orbitals
favor a spin-triplet $S=1$ atomic state because of Hund's coupling in
both \vo\ \cite{mila2000,park2000,dimatteo2002,held2001} and \sroone\
\cite{mravlje2011}.

\begin{figure*}
\includegraphics[width=0.32\columnwidth]{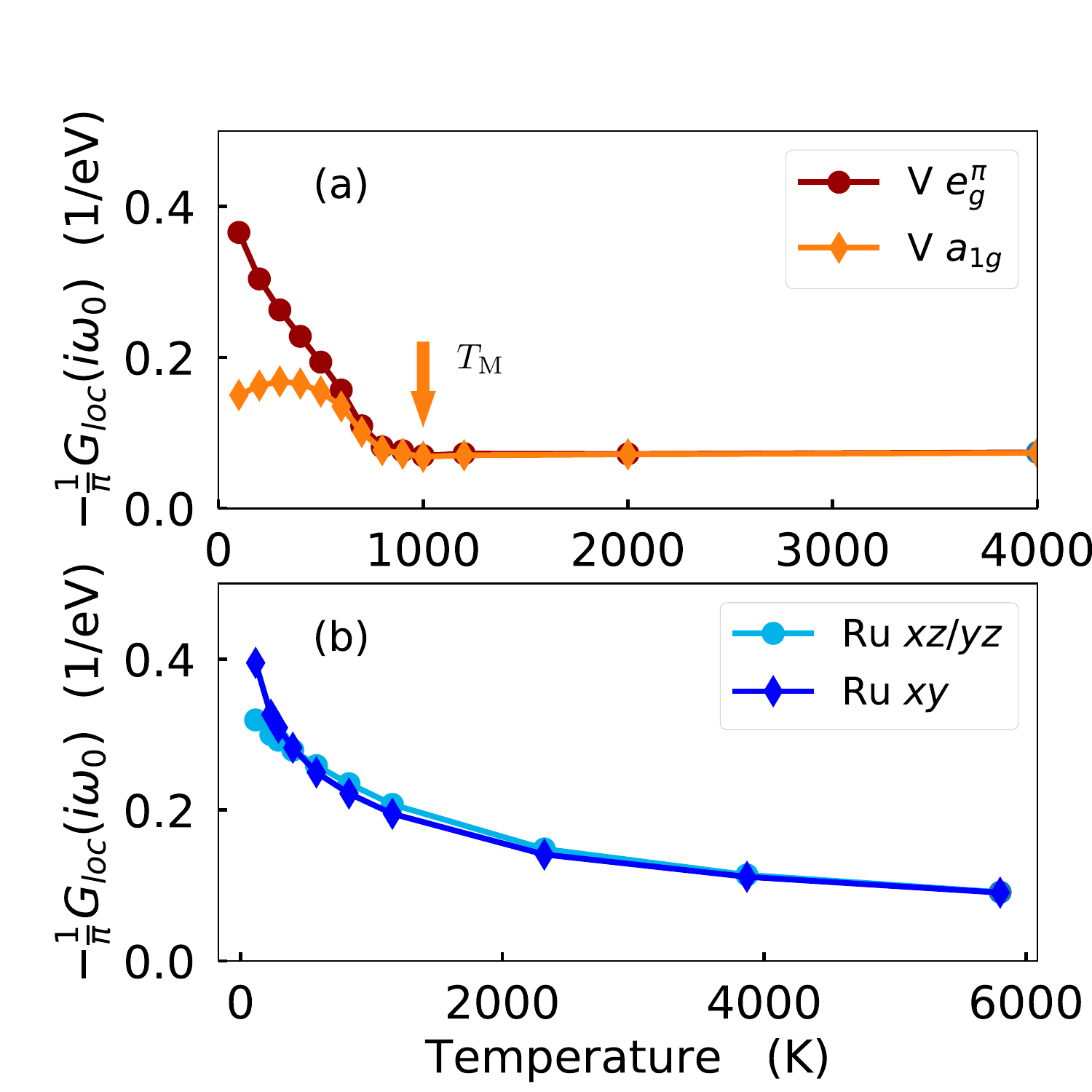}
\includegraphics[width=0.32\columnwidth]{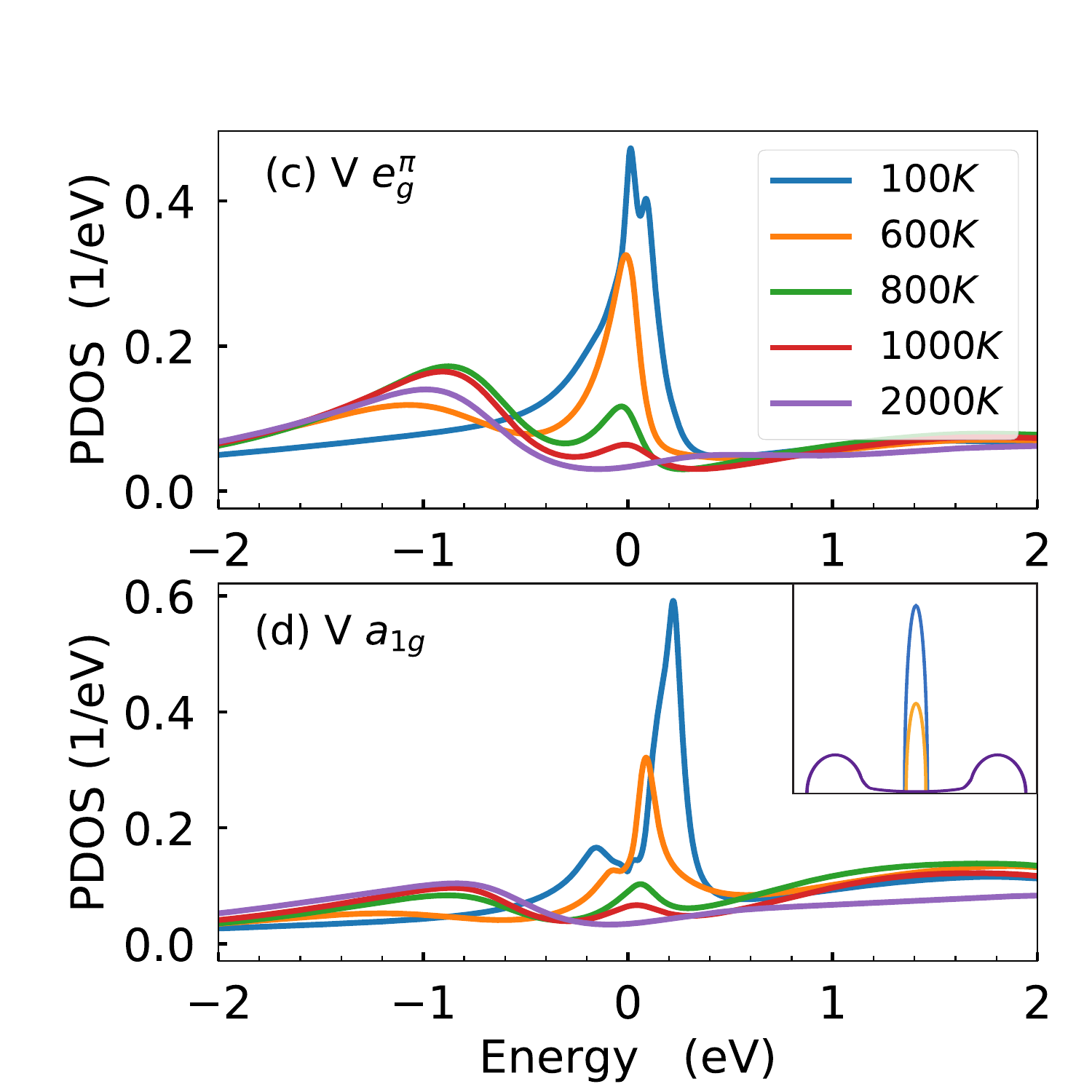}
\includegraphics[width=0.32\columnwidth]{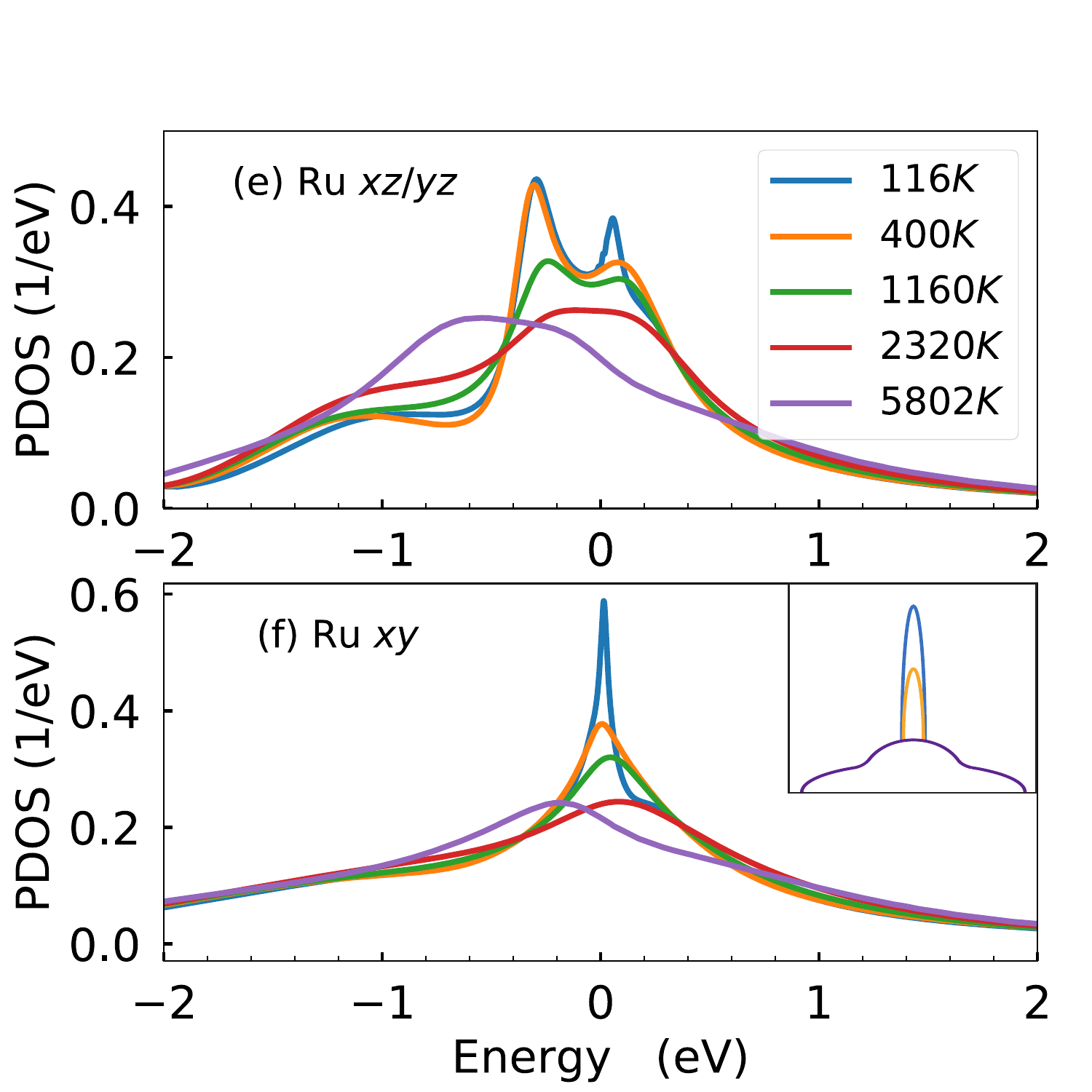}
\caption{The local spectra of the correlated orbitals in
  \vo\ [panels (a,c,d)] and
  \sroone\ [panels (b,e,f)]
  exhibit different behaviors in their
  temperature dependence. (a,b) The density of states at the
  Fermi level, estimated by
  $D(i\omega_0)=-\frac{1}{\pi}{\rm{Im}} G(i\omega_0)$. (c-f) The correlated real-frequency spectra (PDOS),
$D(\omega)=-\frac{1}{\pi}{\rm{Im}} G(\omega)$.
 $D(i\omega_0)$ shows a suppression at a characteristic
  temperature $T_{M}=1000\K$ (indicated by the purple
  arrow) in \vo\ (a), while it evolves smoothly in
  \sroone\ (b). As temperature decreases, in \vo\ the
  coherence resonance of both $e^{\pi}_g$ and $a_{1g}$
  orbitals emerges from the pseudogap regime with low
  density of states between two incoherent peaks (c,d),
  while in \sroone\ the coherence resonance of both
  $d_{xz/yz}$ and $d_{xy}$ orbitals emerges from a single
  broad incoherent peak with large finite density of
  states at the Fermi level (e,f). The insets in (d,f), 
\jvd{repeated from Fig.~\ref{sketch},} 
are cartoons of the temperature dependence of the Mott and Hund PDOS.
}
\label{dos}
\end{figure*}

\JvD{\bheading{Local spectra}}

We compute the spectra of the relevant correlated orbitals in \vo\ and
\sroone\ up to high temperature with DFT+DMFT.  \jvd{We have not taken
  into account the effects of the temperature-dependent changes in lattice parameters, which
  have been shown to be very important in materials near the Mott
  transition such as \vo\ \cite{Baldassarre2008}. Nevertheless, the
  LDA+DMFT calculations here bring a degree of realism, such as band
  structure and crystal field effects, which is not present in the
  3HHM calculations discussed further below.}  We focus first on the
density of states at the Fermi level, estimated via
$D(i\omega_0)=-\frac{1}{\pi}{\rm{Im}} G(i\omega_0)$ ($\omega_0$ is the
first Matsubara frequency, $G$ the computed local Green's function).
Fig.~\ref{dos}(a) depicts the temperature dependence of $D(i\omega_0)$
for $e^{\pi}_g$ and $a_{1g}$ orbitals in \vo. The results show that
both orbitals share a characteristic temperature, $\TM=1000K$:
$D(i\omega_0)$ is fairly flat at temperatures above \TM, which implies
approximately ``rigid", i.e. temperature-independent, spectra. Below
\TM, $D(i\omega_0)$ gradually acquires a larger magnitude in both
orbitals as temperature is lowered, signaling the formation of a
quasiparticle resonance.  We note that it increases monotonically with
decreasing temperature in the $e^{\pi}_g$ orbitals, but in the
$a_{1g}$ orbital it first increases and then decreases a little. Thus
at low temperature the density of states at the Fermi level has a
dominant $e^\pi_g$ character. We emphasize that the evolution of
$D(i\omega_0)$ is smooth and a first-order MIT is not involved. By
contrast, in \sroone\ the temperature dependence of the densities of
states, $D(i\omega_0)$, of $d_{xz/yz}$ and $d_{xy}$ orbitals is very
different, as depicted in Fig.~\ref{dos}(b).  For both orbitals,
\JvD{$D(i \omega_0)$ increases} as temperature is decreased, with gradually
increasing slope, showing no flat regime even at extremely high
temperatures, where their values are already larger than those for
\vo\ above \TM.  In contrast to the case of \vo, a quasiparticle
resonance is present even at the highest temperatures studied and,
thus, no characteristic temperature is found for its onset, as
discussed in the next paragraph.

We also study the correlated real-frequency 
\JvD{projected density of states (PDOS),}
$D(\omega)=-\frac{1}{\pi}{\rm{Im}} G(\omega)$, \JvD{for the different
  orbitals}. These are obtained by analytically continuing the
computed Matsubara self-energy and then computing the local Green's
function.  The results for \vo\ are depicted in Fig.~\ref{dos}(c,d).
At very high temperatures, we observe a typical Mott feature: a
pseudogap exists at the Fermi level, \JvD{between two broad humps in the
  incoherent spectra, with maxima near} $-1\eV$ and $2\eV$. With
decreasing temperature spectral weight is transferred from the
high-energy humps into the pseudogap and a quasiparticle peak emerges
similarly in both orbitals at the Fermi level on top of the
pseudogap. The characteristic temperature for the onset of the
formation of the coherence resonance is roughly consistent with
$\TM=1000\K$ determined above. As temperature decreases further, the
magnitude of the coherence peak in both orbitals increases gradually,
and at very low temperature both orbitals show a coherence resonance
with a pronounced, thin cusp.  In the $e^{\pi}_g$ orbitals the
resonance is peaked at the Fermi level while the $a_{1g}$
quasiparticle peak slightly moves away from the Fermi level \JvD{when
  the temperature is lowered, thus reducing the density of states at
  the Fermi level.  The resulting} temperature evolution of the
zero-frequency density of states in both orbitals is consistent with
the $D(i\omega_0)$ discussed above, including the non-monotonic
behavior of the $a_{1g}$ orbital in Fig.~\ref{dos}(a).  For \sroone\
the slow increase of the density of states at the Fermi level,
$D(i\omega_0)$, with decreasing temperature becomes clear from the
PDOS, shown in Fig.~\ref{dos}(e,f).  The correlated high-temperature
local spectra are characterized by a single broad feature (no
side-humps), which shifts its position slightly towards the Fermi
level \JvD{with decreasing temperature, while its shape remains almost
  unchanged.} This is very different from the spectra in \vo,
\JvD{which at high temperatures show a pseudogap between
two broad side peaks.}
When the temperature is decreased further, a sharp narrow peak
gradually develops in both orbitals from the broad, incoherent
feature. In this process only a small fraction of spectral weight is
transferred from higher frequencies to a 1$\eV$ range around the Fermi
level.  At low temperature, the spectra of both $d_{xz/yz}$ and
$d_{xy}$ orbitals are similar to their corresponding DFT values with a
renormalized bandwidth and show a pronounced, thin cusp as in the case
of \vo.

\begin{figure}
\includegraphics[width=0.7\columnwidth]{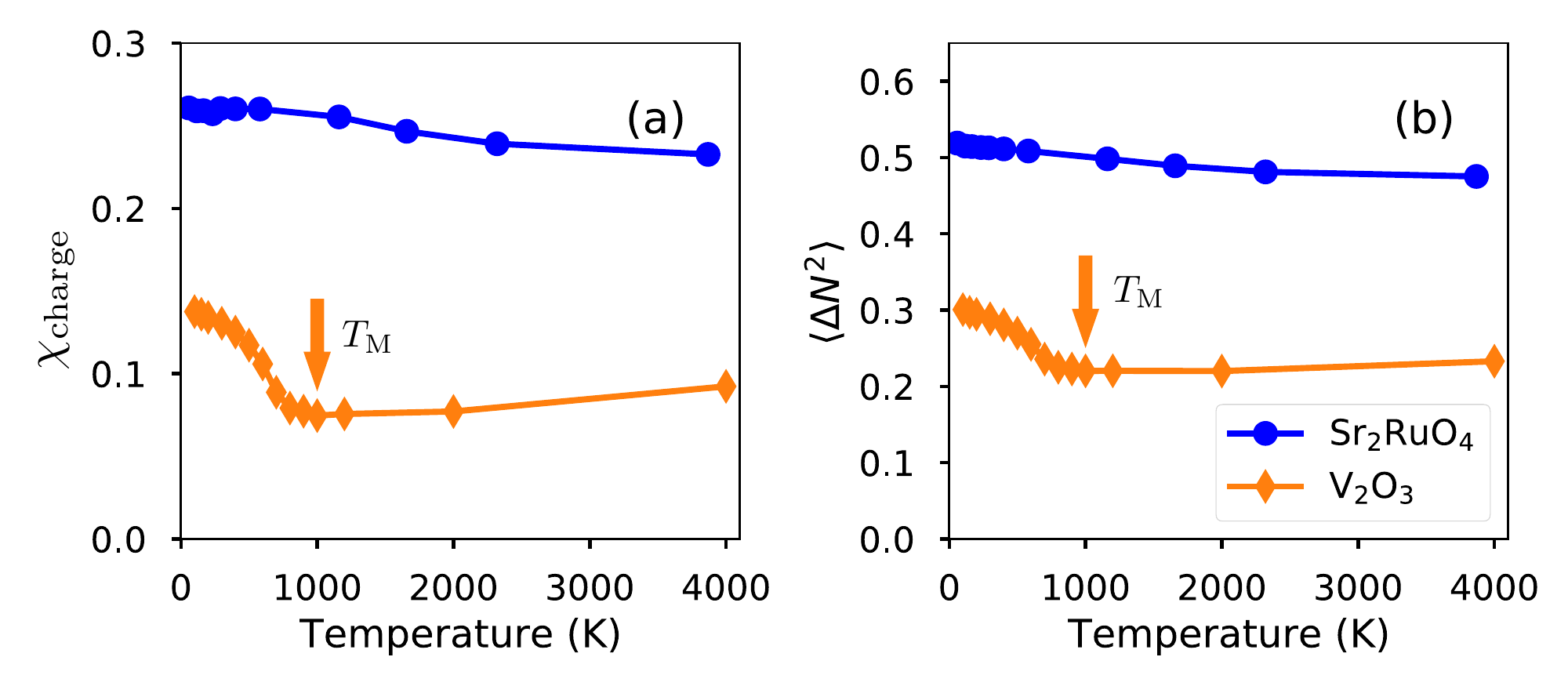}
\caption{(a) The static local charge susceptibility
  $\chi_{\rm{charge}}$, and (b) local charge fluctuation
  $\langle \Delta N^2\rangle$, \JvD{computed for} \vo\ (diamonds) and
  \sroone\ (circles).  \JvD{In the Hund system \sroone\ both
    $\chicharge$ and $\langle \Delta N^2\rangle$ are large and only
    weakly dependent on temperature. By contrast, in the Mott system \vo\
    they} are much smaller and strongly temperature-dependent. The
  purple arrows indicate that in \vo\ the minima of the local charge
  susceptibility and fluctuation occur at the same temperature scale,
  \TM, \JvD{as that marking the emergence of the quasiparticle peak in 
the local PDOS.}}
\label{chg}
\end{figure}
%
\jvd{The different temperature dependences of the local spectra 
of \vo\ and \sroone\ can be viewed as fingerprints distinguishing 
Mott from Hund systems, respectively.}
With decreasing temperature the quasiparticle resonance of \vo\
emerges from a high-temperature pseudogap regime with very low density
of states between incoherent spectra [see purple curve in the cartoon
in inset of Fig.~\ref{dos}(d)]. This is consistent with the widely
held belief that Mott physics governs \vo. It is described by a single
characteristic temperature scale, \TM, which indicates the onset of
formation of the quasiparticle resonance. \JvD{By contrast, for
  \sroone\ the} quasiparticle resonance develops with decreasing
temperature from a single incoherent peak that \JVD{has} a large value
at the Fermi level \JvD{already} at very high temperature [see purple
curve in the cartoon in inset of Fig.~\ref{dos}(f)]. The demonstration
of these two distinct routes towards forming the coherent Fermi-liquid
at low temperature is one of the main results of this work.

\JvD{\bheading{Susceptibilities}}
We next consider the static local charge susceptibility,
$\chi_{\rm{charge}} =\int_0^\beta \langle N_d(\tau)N_d(0)\rangle
d\tau-\beta\langle N_d\rangle^2$,
\JvD{and the local charge fluctuations,
  $\langle \Delta N^2\rangle=\langle N_d^2\rangle -\langle
  N_d\rangle^2$,
  shown in Figs.~\ref{chg}(a) and \ref{chg}(b), respectively.  ($N_d$
  is the total occupancy of $t_{2g}$ orbitals.)  For both materials,
  the behavior of $\chicharge$ mimics that of
  $\langle \Delta N^2\rangle$, hence we focus on the latter below.
  $\langle \Delta N^2\rangle$ is much smaller, \JVD{with} a much
  stronger temperature dependence, in \vo\ than in \sroone.  For \vo,
  $\langle \Delta N^2\rangle$ initially remains small and almost
  constant with decreasing temperature, signifying the suppression of
  charge fluctuations in the pseudogap regime. It then increases
  rather abruptly, signifying the onset of charge delocalization, at
  the same temperature, $T_M = 1000\mathrm{K}$ (purple arrow in
  Fig.~\ref{chg}(a)), as that where the quasiparticle peak begins to emerge.
By contrast,
for \sroone\ $\langle \Delta N^2\rangle$ 
exhibits only a weak temperature dependence, persisting  up
to the highest temperature studied but  changing by less than $10\%$ over
this range.} 

\begin{figure}
\includegraphics[width=0.42\columnwidth]{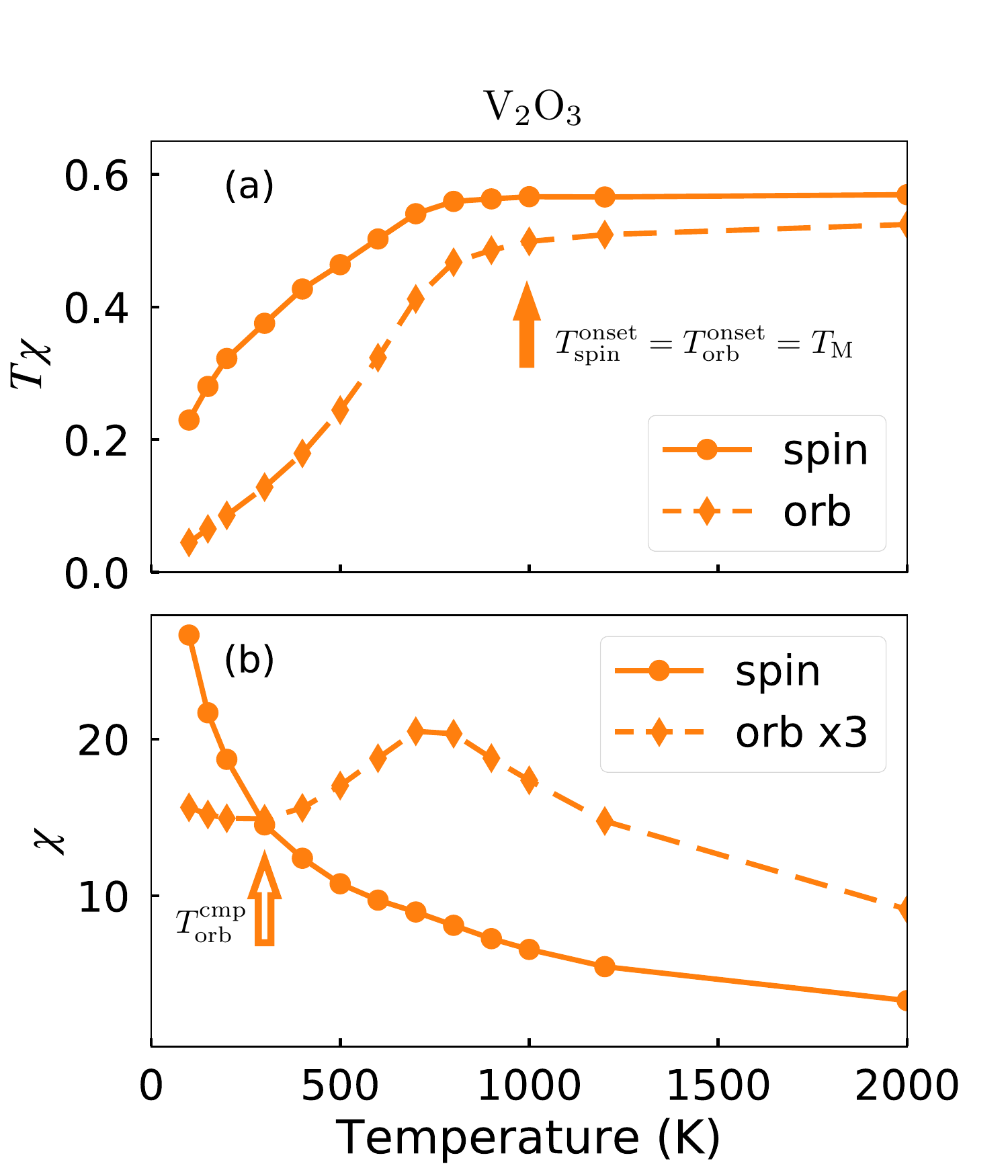}
\includegraphics[width=0.42\columnwidth]{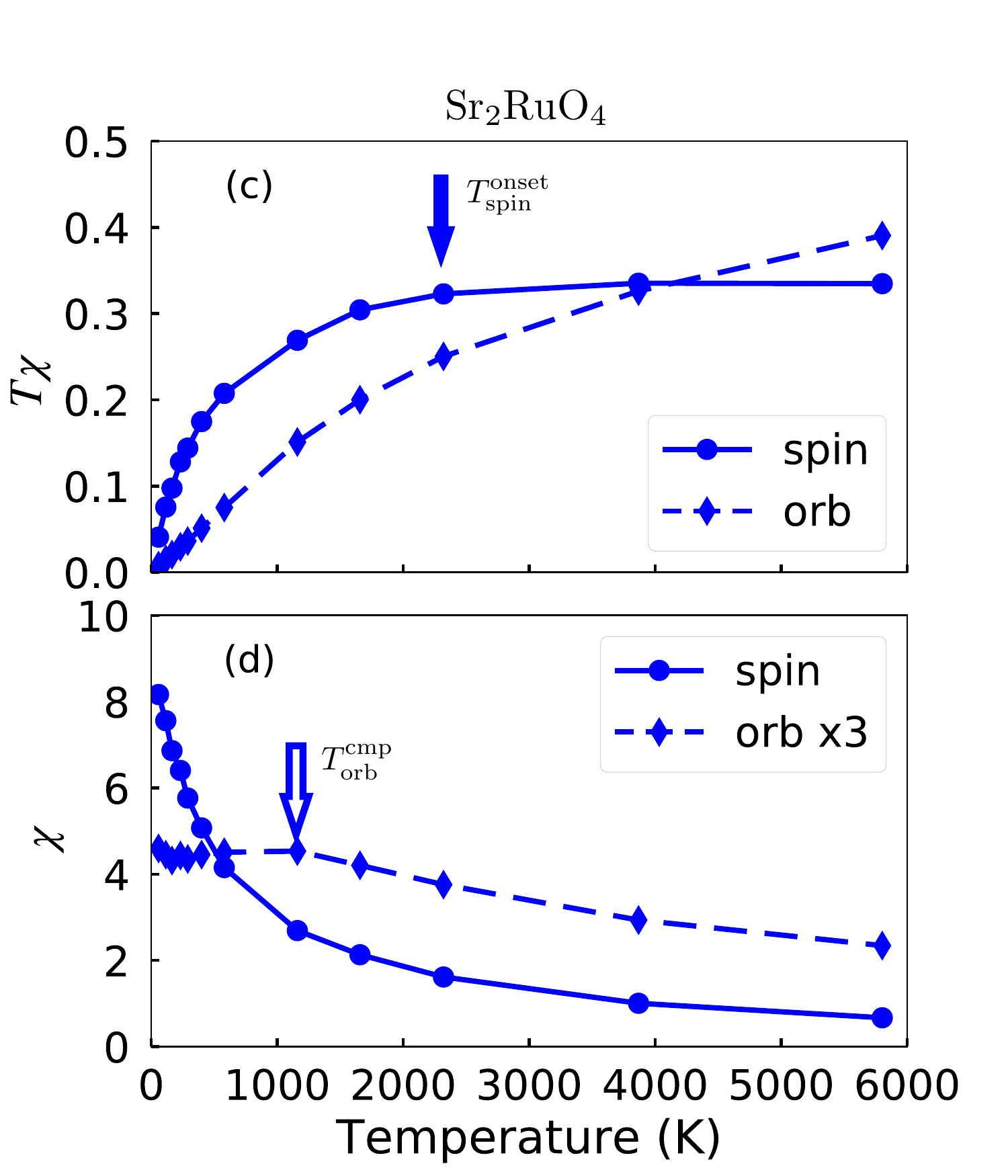}
\caption{The static local orbital and spin susceptibilities $\chiorb$
  and $\chispin$ of \vo\ (a,b) and \sroone\ (c,d), \JvD{plotted as
  functions of temperature, with $T\chi$ and $\chi$ shown in the upper
  and lower panels, respectively}.  The Curie law holds above $1000\K$
  (indicated by open purple arrow) in the spin and orbital
  susceptibility of \vo\ (a). In \sroone\ the spin susceptibility
  follows Curie-like behavior above $2300\K$ (indicated by open cyan
  arrow), while the orbital susceptibility does not follow a Curie law
  in the temperature range studied (c). The spin susceptibility of
  both materials (b,d) does not saturate at the lowest accessible
  temperature, indicating an even lower Fermi-liquid scale. The
  orbital susceptibility has only weak temperature dependence below
  $300\K$ in \vo\ (b) and below $1100\K$ in \sroone\ (d) (indicated by
  filled blue arrows).}
\label{sus}
\end{figure}
We have also computed the static local spin and orbital
susceptibilities, defined as
$\chispin=\int_0^\beta \langle S_z(\tau) S_z(0) \rangle d \tau$ and
$\chi_{\rm{orb}}=\int_0^\beta \langle \Delta N_{\rm{orb}}(\tau) \Delta
N_{\rm{orb}} \rangle-\beta \langle \Delta N_{\rm{orb}} \rangle^2$,
where $S_z$ is the total spin momentum in the $t_{2g}$ orbitals,
$\Delta N_{\rm{orb}}=N_{a}/2-N_{b}$ is the occupancy difference per
orbital, and $(a,b)$ denotes $(e^{\pi}_g,a_{1g})$ in \vo\ and
$(xz/yz,xy)$ in \sroone, respectively. The results are depicted in
Fig.~\ref{sus}.  In \vo, both the spin and orbital susceptibilities
exhibit Curie behavior, i.e. $T\chispin$ and $T\chi_{\rm{orb}}$ are
approximately constant at high temperature
(Fig.~\ref{sus}(a)). Notably, with decreasing temperature deviations
from the Curie behavior set in at the same characteristic temperature,
$\TM=1000\K$, as that determined above from the local PDOS
evolution. Thus, spin and orbital degrees of freedom start to be
screened simultaneously with the formation of a coherence resonance in
the prototype Mott system \vo, $\TorbO= \TspinO=\TM$. By contrast, in
the Hund material \sroone, Curie-like behavior in the spin
susceptibility is seen only at very high temperatures. With decreasing
temperature, it ceases already at around $\TspinO\simeq 2300\K$
(Fig.~\ref{sus}(c)), a scale much higher than that in \vo. For the
orbital susceptibility the situation is even more extreme: it does not
show Curie behavior even at the highest temperature studied
($\TorbO\geq 6000\K$).  This is evidence of spin-orbital separation in
\sroone: the screening of the orbital degrees of freedom starts at
much higher temperature than that of the spin degrees of freedom,
$\TorbO\gg \TspinO$.  Hence the onset of deviations from Curie-like
behavior in the spin/orbital susceptibility, i.e.~the onset of
screening of spin/orbital degrees of freedom, is very different in
\vo\ and \sroone. \JvD{These differences constitute another
set of fingerprints distinguishing  Mott from Hund systems.}
It will be further analyzed below in the
context of our 3HHM calculations.

\JvD{Next we discuss the completion of orbital and spin screening,
  characterized by the temperature scales, $\TorbC$ and $\TspinC$,
  below which the corresponding susceptibilities become constant. 
  In both materials, $\chiorb$ 
\ks{seems to become} essentially constant
 at low temperatures, with the orbital screening completion scale in 
\vo,
  $\TorbC\simeq 300\K$ (Fig.~\ref{sus}(b)), being much smaller than in
  \sroone, $\TorbC\simeq 1100\K$~(Fig.~\ref{sus}(d)).} By contrast, in
  both materials the spin susceptibility increases with decreasing
  temperature, and is not fully screened even at the lowest
  temperature studied. This is consistent with the experimental
  observations that in both materials $\TFL$ is as low as about
  $25\K$, and $\TFL$ provides an estimation for \TspinC at which the
  spin degrees of freedom are fully screened.  
  \ks{In summary, we clearly deduce spin-orbital separation in the
    completion of screening, $\TorbC\gg \TspinC$, for the Hund metal
    \sroone, while this effect is less pronounced in the Mott material
   \vo, where $\chiorb$ in addition shows a bumb before it tends
    to saturate at lower temperatures.}

In Hund metals the spin-orbital separation has
  been pointed out in numerical studies of the frequency dependence of
  the local self-energy and susceptibilities
  \cite{yin2012,stadler2015,stadler2018} and in an analytical estimate
  of the Kondo scales \cite{aron2015}.  Here, our results reveal that
  it also occurs in the temperature domain.  
  We note that our computed spin susceptibility of \sroone\
  is consistent with earlier results using a narrower temperature range
  \cite{mravlje2015}.

\JvD{\bheading{Entropy}}
In both materials the entropy of the correlated atom reaches a plateau
of $\ln(3\times 3 =9)$ at high temperatures as expected for a high
spin ($S = 1$) state with large contribution from three active
$t_{2g}$ orbital degrees of freedom \cite{georges2013,stadler2015}.
Notably, in Mott systems, with decreasing temperature the plateau
persists down to the temperature scale, \TM, until which both the spin
and orbital degrees of freedom remain unquenched and the quasiparticle
resonance has not yet formed in the pseudogap of the local correlated
spectrum. These results are discussed in the supplement \cite{note_supp}.

\JvD{\aheading{Three-band Hubbard-Hund model}
\smallskip

\bheading{Model Hamiltonian}
We now turn to the 3HHM, described by the  
Hamiltonian}
\begin{eqnarray}
&& \hat{H}  =  \sum_{i} \left(  - \mu \hat N_{i} 
+ \hat{H}_\interact [\hat d^\dag_{i\nu}] \right) 
+\sum_{\langle ij\rangle \nu} t\,
  \hat{d}^{\dagger}_{i\nu}\hat{d}^{\phantom{\dagger}}_{j\nu} ,
\label{eq:HU} \\
&& \hat{H}_\interact[\hat d^\dag_{i\nu} ] 
=  \tfrac{1}{2}\left( U-\tfrac{3}{2}J \right) 
\hat N_i (\hat N_i -1)-{J}\hat{\mathbf S}_i^2
+ \tfrac{3}{4} {J}  \hat N_i .
\text{}\notag 
\end{eqnarray}
The on-site interaction term incorporates Mott and Hund physics
through $U$ and $J$ respectively. $\hat d^\dagger_{i \nu}$ creates an
electron on site $i$ of flavor $\nu = (m\sigma)$, which is composed of
a spin ($\sigma \! = \uparrow,\downarrow$) and orbital ($m=1,2,3$)
index.  $\hat n_{i\nu} =\hat{d}^{\dagger}_{i\nu}\hat{d}^\pdag_{i\nu}$
counts the electrons of flavor $\nu$ on site $i$.
$\hat N_i =\sum_{\nu}\hat n_{i\nu}$ is the total number operator for
site $i$ and $\hat{\mathbf S}_i$ its total spin, with components
$\hat S_i^\alpha = \sum_{m\sigma\sigma'}\hat{d}^{\dagger}_{i m\sigma}
\tfrac{1}{2}\sigma^\alpha_{\sigma\sigma'}\hat{d}_{i m\sigma'}$,
where $\sigma^{\alpha}$ are Pauli matrices.  We take a uniform hopping
amplitude, $t=1$, serving as energy unit in the 3HHM, and a Bethe
lattice in the limit of large lattice coordination. The total width of
each of the degenerate bands is $W=4$. We choose the chemical
potential $\mu$ such that the total filling per lattice site is
$\langle N_i \rangle=2$. The model is solved numerically exactly using
DMFT+NRG 
\cite{stadler2015,stadler2018}.

\begin{figure}[tbph]
\includegraphics[width=1\columnwidth,trim=0mm 23mm 0mm 17mm]{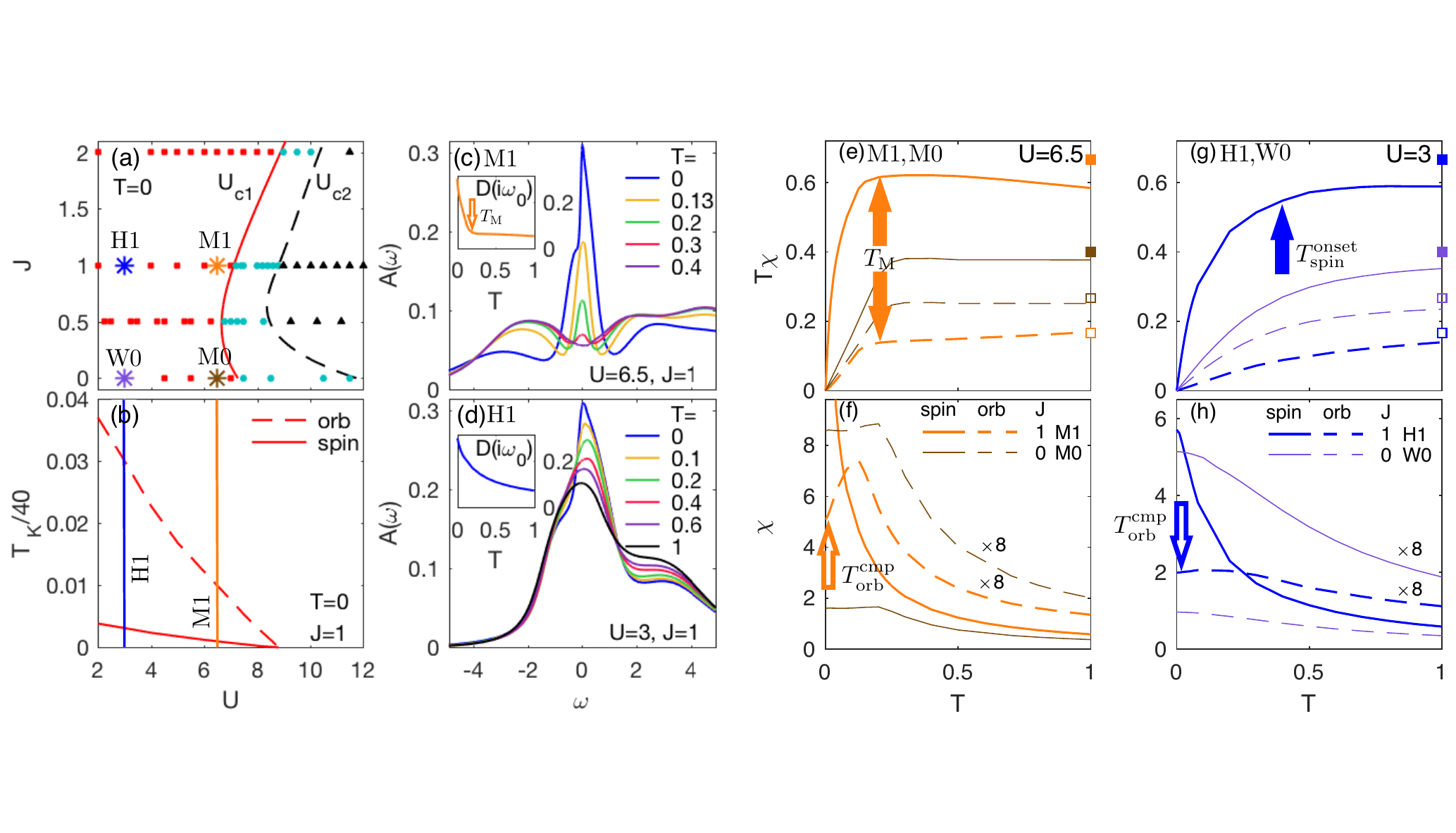}
   \renewcommand{\baselinestretch}{1}
   \caption{Disentangling features of Mott and Hund physics in a
     DMFT+NRG study of the
     3HHM 
     (unit of energy: hopping amplitude $t = 1$).  (a)
     The $T=0$ phase diagram \JVD{(cf.\ Fig.~5 of
       \cite{stadler2018})} reveals three phases in the $J$-$U$-plane:
     a metallic phase (red squares), a coexistence region (blue
     circles), and an insulating phase (black triangles), separated by
     two phase transition lines $U_{c1}$ (solid red curve), and
     $U_{c2}$ (dashed black curve), respectively.  \JvD{In panels
       (c-h) we focus on \jvd{four} parameter combinations, indicated
       in (a) by colored asterisks: two Mott systems with $U=6.5$ near
       the $U_{c1}$ phase transition line, with $J=1$ (M1) \jvd{or
         $J=0$ (M0)}; and two systems with $U=3$ far from the
       transition and deep in the metallic state, a Hund system with
       $J=1$ (H1) \jvd{and a weakly correlated system with $J=0$
         (W0)}.}  (b) The Kondo temperatures, here shown for $J=1$
 \JVD{(cf.~Fig.~12 of \cite{stadler2018})} 
     are extracted from frequency-dependent susceptibilities at $T=0$,
     as defined in \cite{stadler2015}. \TKspin (\TKorb) corresponds to
     the screening of spin (orbital) degrees of freedom.  Fermi liquid
     behavior sets in below the temperature scale
     $\TKspin/40\approx \TspinC$.  \JvD{Orange and blue} vertical
     lines mark the values of $U$ used for M1 and H1, respectively.
     (c,d) The temperature dependence of the LDOS for M1 and H1,
     respectively. The energy scale of the lowest bare atomic
     excitations, $\pm \Eatomic = \pm (\frac{1}{2}U-J)$, and thus the
     Hubbard bands, is much larger for M1 than H1. For M1 in (c) a
     pseudogap (a typical Mott feature) emerges \JvD{when the
       temperature increases past a characteristic value, $T_M$, which
       lies far} below the rather large scale
     $\Eatomic^{\rm M1} \simeq 2.25$.  By contrast, for H1 in (d) a
     pronounced peak in the density of states still exists even at
     very high temperatures, $T>0.5$, that exceed the rather small
     scale $\Eatomic^{\rm H1} \simeq 0.5$.  The insets of (c,d) show
     the LDOS at the Fermi level, estimated by
     $D(i\omega_0)=-\frac{1}{\pi}{\rm{Im}}G(i\omega_0)$, for M1
     (orange) and H1 (blue).  (f-h) The local spin and orbital
     susceptibilities are shown as functions of $T$ for M1 (orange)
     \jvd{and M0 (brown)} in (e,f), and for H1 (blue) and \jvd{W0
       (purple)} in (g,h), with $T\chi$ depicted in the upper panels
     (e,g) and $\chi$ in the lower panels (f,h).  (e,g) For
     temperatures well above $\TorbO$ or $\TspinO$, respectively,
     $T\chiorb$ and $T \chispin$ approach plateaus, indicative of a
     Curie law, as expected for unscreened spin or orbital degrees of
     freedom. The observed plateau heights are roughly comparable to
     the values expected \cite{Hanl2014b} for free local moments
     \jvd{with occupancy strictly equal to 2 and, for M1 and H1 (M0
       and W0), spin equal to 1 (and 0),
       for which
       $T\chi^\free_\spin = \tfrac{1}{3} \langle {\bf \hat S^2}
       \rangle = 2/3\,(2/5)$
       and
       $T \chi^\free_\orb = \tfrac{1}{8} \langle {\bf \hat T^2}\rangle
       = 1/6\,(4/15)$}
     indicated by filled and empty squares on the right vertical axes,
     respectively.  (Deviations of the observed plateaus from these
     local moment values reflect admixtures of states with different
     occupancy or spin.)  For M1 in (e), the Curie law ceases to hold
     for both $\chiorb$ and $\chispin$ below about $\TM\simeq0.2$
     (orange arrow).  For H1 in (g), $\chispin$ deviates from a
     Curie-like behavior below about $T\simeq0.4$ (blue arrow), while
     $\chiorb$ does not follow a Curie law in the temperature range
     displayed.  \jvd{For M0 in (e) and W0 in (g), deviations from
       Curie behavior set in at similar temperatures for $\chispin$
       and $\chiorb$.}  \JvD{Thus the onset of screening shows
       spin-orbital separation for H1, but not for M1 (due to its
       proximity to the Mott transition),} \jvd{and also not for M0
       and W0 (since these have $J=0$).}  (f,h) For both M1 and H1,
     $\chispin$ saturates at very low Fermi-liquid temperatures (not
     displayed here, but clearly deducible from the underlying
     zero-temperature NRG data \cite{stadler2015,stadler2018}).  By
     contrast, $\chiorb$ is approximately temperature independent
     below $T=0.01$ (orange arrow) for M1 in (f) and below $T=0.03$
     (blue arrow) for H1 in (h).  \jvd{For M0 in (f) and W0 in (h),
       $\chiorb$ and $\chispin$ become temperature independent at
       similar temperatures.}  \JvD{Thus, the completion of screening
       shows \ks{tendencies of} spin-orbital separation for M1 and H1 (since $J\neq 0$),}
     \jvd{but not for M0 and W0 (since $J=0$). Moreover, $ \TspinC$
       and hence $\TFL$ is much smaller for $J \neq 0$ than for
       $J=0$.}  }
\vspace{-.2in}
\label{nrg}
\end{figure}

\JvD{\bheading{Phase diagram}}
The 3HHM 
enables the exploration of a broad
region of parameters at arbitrary low temperatures. Fig.~\ref{nrg}(a)
illustrates the $J$-$U$ phase diagram at $ T=0$. 
\jvd{To illustrate the difference between large and small $U$, and
  non-zero and zero $J$, we will focus on four parameter combinations,
denoted by M1, H1, M0 and W0, depicted by asterisks in Fig.~\ref{nrg}(a),
and defined in detail in the figure caption. 
The Mott system M1 ($U=6.5$) 
and the Hund system H1 ($U=3$), both with $J=1$, lie close to
or far from the Mott transition and qualitatively mimic} \vo\ and
\sroone\ respectively, considering their multi-orbital nature, sizable
Hund's coupling and  distances to the Mott
transition. \jvd{The Mott system M0 ($U=6.5$) and the weakly correlated system
  W0 ($U=3$) , both with $J=0$, illustrate the consequences of turning off
  Hund's coupling altogether.} 

Fig.~\ref{nrg} displays the LDOS, 
$A(\omega)=-(1/\pi) {{\rm{Im}}}G(\omega)$, for M1 and H1 (c,d);
the corresponding density of states
at the Fermi level (insets of (c,d)), estimated by
$D(i\omega_0)=-\frac{1}{\pi}\mathrm{Im}G(i\omega_0)$; and the static local
susceptibilities $T \chi $ (e,g) and
$\chi\equiv\chi_{\rm{d}}(\omega=0)$ (f,h) for the spin (solid) and orbital
(dashed) degrees of freedom of M1 (orange), \jvd{M0 (brown)}, 
H1 (blue) and \jvd{H0 (purple)}. 
The corresponding dynamical real-frequency spin
and orbital susceptibilities are defined as
$\chi_{\rm{d},\spin}(\omega) = \tfrac{1}{3}\sum_\alpha \langle \hat
S^\alpha \mbox{$\parallel$} \hat S^\alpha \rangle_\omega$ and
$\chi_{\rm{d},\orb}(\omega) = \tfrac{1}{8} \sum_a \langle \hat
T^a\mbox{$\parallel$} \hat T^a \rangle_\omega$, respectively
\cite{Hanl2014b,weichselbaum2012b}, where
$\hat T^a = \sum_{mm'\sigma}\hat{d}^{\dagger}_{m\sigma}
\tfrac{1}{2}\tau^a_{mm'}\hat{d}_{m'\sigma}$
are the impurity orbital operators with the SU(3) Gell-Mann matrices,
$\tau^a$, normalized as ${\rm Tr}[ \tau^a\tau^b ] =2\delta_{ab}$.  The
qualitative similarities (especially of Fig.~\ref{nrg}(c,d,g,f,h))
with those in Fig.~\ref{dos} and Fig.~\ref{sus} are obvious, in spite
of the simplified band structure and the absence of crystal fields.
\jvd{Let us now discuss these in detail.}

\JvD{\bheading{Mott system M1}}
The basic features of \vo\ are reproduced \jvd{by M1, lying}
close to the phase transition line. At high temperatures we also observe a
pseudogap in the incoherent spectra at the Fermi level, formed between
two broad Hubbard sidebands, one at negative and one with minor
substructure at positive frequencies, respectively (Fig.~\ref{nrg}(c),
red and purple curves). With decreasing temperature spectral weight is
transferred from these high-energy humps into the pseudogap, building
up a clear peak at about $\TM\simeq0.2$, which evolves into a
pronounced, sharp coherence resonance at very low temperature
(Fig.~\ref{nrg}(c), blue curve]. This behavior is confirmed by
$D(i\omega_0)$ (inset of Fig.~\ref{nrg}(c)).  $T \chi$ shows flat
Curie behavior for both orbital and spin degrees of freedom in the
pseudogapped phase at high temperatures. With decreasing temperature
the orbitals and spins start to get screened simultaneously at the
same energy scale, $\TorbO=\TspinO=\TM$, at which the resonance
emerges in the pseudogap (Fig.~\ref{nrg}(e)), analogously to the
behavior in the Mott material \vo.

\JvD{\bheading{Hund system H1}} For the Hund system H1 far from the
phase transition line, the physical properties (LDOS, local spin- and
orbital susceptibility) show the same qualitatively behavior as for
\sroone.  At very high temperatures (above
$T^{\rm{onset}}_{\rm{spin}}\simeq0.4$, red, purple and black curves in
Fig.~\ref{nrg}(d)) the local spectral function has a large density of
states \JvD{near the Fermi energy}, in contrast to the pseudogap
present for M1.  At these large temperatures the spin susceptibility
shows Curie-like behavior, in that $T\chispin $ is essentially
constant there, whereas $T \chi_{\orb}$ decreases with decreasing
temperature (Fig.~\ref{nrg}(g)). This indicates that the spins are
still large and (quasi-)free while the orbitals are already being
screened, $\TspinO\ll\TorbO$.  Below $\TspinO$ also the spin degrees
of freedom get screened and a pronounced quasiparticle peak gradually
develops, with a sharp cusp at low frequencies and very low
temperatures (blue curve in Fig.~\ref{nrg}(d)). In contrast \JVD{to M1,} 
$D(i\omega_0)$ \JVD{for H1} is large already at high temperatures and
increases continuously with decreasing temperature (inset of
Fig.~\ref{nrg}(d)).

\JvD{\bheading{Completion of screening}} \ks{In principle,} \JvD{for
  both M1 and H1, i.e.}\ both close to and far from the Mott
transition, orbital screening is completed at a higher temperature
than spin screening. Indeed, an approximately temperature-independent,
Pauli-like susceptibility, is observed for $\chiorb$ (dashed lines)
below $\TorbC$ (indicated by arrows in Figs.~\ref{nrg}(f,h)), while
$\chispin$ (solid lines) still increase with decreasing temperature,
hence $\TorbC\gg \TspinC$. \ks{(This effect is more pronounced for H1,
  i.e. far from the Mott transition.)}  \jvd{By contrast, the
  corresponding M0 and W0 curves in Figs.~\ref{nrg}(f,h), having
  $J=0$, show \ks{no spin-orbital separation for the completion of
    screening, i.e.} $\TorbC\simeq \TspinC$, as described in more
  detail in the figure caption.} 

\JvD{\aheading{Discussion}}

\noindent 
The DMFT solution of the 3HHM \JvD{with nonzero $J$} enables us to
understand the interplay between Mott and Hund physics and its
materials manifestations from an impurity model perspective.  Far from
the transition, a picture in terms of a multi-orbital Kondo model in a
broad-bandwidth metallic bath applies. Standard analysis of the
logarithmic Kondo singularties showed that $\TspinO \ll \TorbO$
\cite{aron2015}. As we approach the Mott boundary, charge fluctuations
are blocked, resulting in well-separated Hubbard bands. Here the onset
of the Kondo resonance is not signaled by logarithmic singularities
but instead it is driven by the DMFT-self-consistency condition
\cite{fisher}.  In this regime the onsets of screening for spin and
orbital degrees of freedom occur at the same scale, \jvd{namely that
  where charge delocalization sets in}.  The spin-orbital separation
in the completion of screening, which occurs at \textit{low
  temperatures} in both Mott and Hund systems, can be understood from
a zero-temperature analysis of the 3HHM.  We define characteristic
Kondo scales \TKorb and \TKspin, from the maximum in the
zero-temperature, frequency-dependent local orbital and spin
susceptibilities \cite{stadler2015}, respectively, and display them in
Fig.~\ref{nrg}(b) for $J=1$ as a function of $U$. We find that
$\TKspin \ll \TKorb$, so an intermediate region with free spins and
quenched orbitals is a generic feature of multi-orbital systems with
significant Hund's coupling as was surmised from earlier studies.  For
both the Hund and Mott system results we deduce that
$T_{\rm K}/40\approx T^{\rm cmp}$: spin-orbital separation in
frequency space thus has a direct manifestation in the completion of
screening as a function of temperature.  Below the spin completion
scale we have a Femi liquid.  As we approach the Mott boundary for
increasing $U$ the spin-orbital separation region shrinks, and all the
energy scales are reduced, as shown in Fig.~\ref{nrg}(b), elucidating
the reduced $\TorbC$ in the Mott system (and in \vo) compared to the
Hund system (and \sroone).

\jvd{Finally, we note that the systems M0 and W0, with $J=0$, show no
spin-orbit separation for the onset or completion of screening (see
 Figs.~\ref{nrg}(e-l) and their discussion in the figure caption). 
Conversely, turning on $J$ pushes the Fermi liquid scale
$\TFL = \TspinC$ strongly downward relative to $\TorbO$.
This significantly reduces the quasiparticle weight
$Z = m/m^\ast$ (which is proportional to $\TFL$ \cite{stadler2018})
and increases the strength of correlations. Hence the Hund system H1
is much more strongly correlated than W0, although $U$ is the
same for both. These differences leave clear fingerprints in 
photoemission spectra, where $Z$ characterizes the slope of the
quasiparticle dispersion. Moreover, the shape of the
quasiparticle peak shows substructure indicative
of spin-orbital separation for sizable $J$, but not for $J=0$.
For a detailed illustration of these points,
see Ref.~\cite{note_supp}, Fig.~S-2. }

\aheading{Conclusions}

\noindent
In conclusion, we revealed contrasting signatures of Mottness and
Hundness in two archetypal materials, \vo\ and \sroone, in the
formation of the quasiparticle resonance in the local correlated
spectra, and in the temperature dependence of the charge, spin, and
orbital susceptibility as well as the impurity entropy. Mott and Hund
physics manifest in the process in which the atomic degrees of freedom
at high energies evolve towards low energies to form fermionic
quasiparticles. We highlight the observation of four temperature
scales that characterize the onset and the completion of screening of
the spin and the orbital degrees of freedom. 
\ks{We find that a non-zero Hund's coupling leads to  
spin-orbital separation in the completion of screening at low temperatures,
$\TorbC \gg \TspinC$, and this is more pronounced for  Hund systems.}
However, Mott and Hund systems
show contrasting behavior at intermediate to high energies, due to the
very different relations of their overall quasiparticle peak width and
their atomic excitation scales: $\TorbO \ll \Eatomic$ for Mott systems
vs.  $ \Eatomic \lesssim \TorbO $ for Hund systems.  In the Mott
system \vo\ the strong Coulomb repulsion localizes the charge at high
temperature, with decreasing temperature the onset of \JvD{charge
  localization triggers the simultaneous onset of} the screening of
the spin and orbital degrees, accompanied by the formation of the
coherence resonance at $\TM \equiv \TspinO=\TorbO \ll \Eatomic$.  In
contrast, in \sroone\ Coulomb repulsion is much weaker, so that no
charge localization occurs even at very high temperatures. Therefore,
charge fluctuations triggering the onset of screening are possible
\JvD{even at} high temperatures, leading -- due to the presence of
sizeable Hund's coupling -- to a clear separation in the energy scales
at which this screening sets in for spin and orbital fluctuations,
with $\TspinO \ll \TorbO$.  All these findings are generic and do not
depend on microscopic details. They only require a sizeable Hund's
coupling, and are controlled by the distance to the Mott localization
boundary.  This is confirmed by a DMFT+NRG study of a model 3-band
Hubbard-Hund Hamiltonian, thus establishing a general phenomenology of
Mottness and Hundness in multi-orbital systems.  Our results give not
only new perspectives into the archetypical strongly correlated
materials, \vo\ and \sroone, but will be useful in interpreting
experimental measurements on other correlated metals and in
identifying the origin of their correlations.

\aheading{Methods}

\noindent 
The two prototype materials are investigated using
the all-electron DMFT method as implemented in Ref. \cite{haule2010}
based on the WIEN2k package \cite{blaha2001} and the continuous-time
quantum Monte-Carlo (CTQMC) impurity
solver \cite{werner2006a,haule2007}. We used projectors within a large
(20$\eV$) energy window, i.e.~we used a high energy cutoff scale, to
construct local orbitals, thus the oxygen orbitals hybridizing with
the $d$ orbitals were explicitly included. With such a large energy
window the resulting $d$ orbitals are very localized.  In our two
example materials these are the $t_{2g}$ levels of Ru and V atoms,
which we treated dynamically with DMFT, all other states were treated
statically and no states were eliminated in the calculations.  The
nominal ``double counting" scheme with the form
$\Sigma_{DC}=U(n_{\rm imp}-1/2)-\frac{1}{2}J(n_{\rm imp}-1)$ was used where
$n_{\rm imp}$ is the nominal occupancy of $d$ orbitals. The onsite
interactions in terms of Coulomb interaction $U$ and Hund's coupling $J$
were chosen to be $(U,J)=(6.0,0.8)\eV$ for V in \vo\ and
$(U,J)=(4.5,1.0)\eV$ for Ru in \sroone.  
The impurity entropy was
computed by integrating the impurity internal energy up
to high temperature, following   Ref. \cite{haule2015}.
Our DFT+DMFT setup was successful in describing the correlation
effects in both materials \cite{deng2014,deng2016}. 
It captures the 
phase diagram of \vo\ which exhibits a  Mott  MIT   and our computed electronic
structure is consistent with experimental measurements \cite{deng2014}.  The approach also 
describes the electronic structure of
\sroone\ and is in good  agreement with  the results of experimental
measurements \cite{deng2016} and  other DFT+DMFT
calculations \cite{mravlje2011, dang2014, dang2015}.  In addition our
studies \cite{deng2014,deng2016} correctly characterize  the transport and optical
properties of both materials. These successes gave us confidence to  extend
our studies  to even higher temperatures, and for quantities which have 
yet to be measured experimentally. 

We solved the 3HHM 
using DMFT \cite{georges1996} in combination with an efficient multi-band
impurity solver \cite{stadler2015,stadler2018}, the full-density-matrix (fdm)
NRG \cite{weichselbaum2007}.
Our fdmNRG solver employs a complete basis
set \cite{asbasisprl,asbasisprb}, constructed from the discarded
states of all NRG iterations.  Spectral functions for the discretized
model are given from the Lehmann representation as a sum of poles, and
can be calculated accurately directly on the real axis in sum-rule
conserving fashion \cite{peters2006} at zero or arbitrary finite
temperature.  Continuous spectra are obtained by broadening the
discrete data with a standard log-gaussian Kernel of
frequency-dependent \cite{bulla2008,weichselbaum2007} width.  Further,
fdmNRG is implemented in the unified tensor representation of the
QSpace approach \cite{weichselbaum2012b} that allows us to exploit
Abelian and non-Abelian symmetries on a generic level [here
$\text{U(1)}_{\rm{charge}}\times%
\text{SU(2)}_{\rm{spin}} \times\text{SU(3)}_{\rm{orb}}$].  For further
details of our DMFT+NRG calculations see the Supplementary material
of \cite{stadler2015}.

\textbf{Acknowledgments}: Work by X.D. was supported by NSF
DMR-1733071. Work by K.H. was supported by NSF DMR 1405303. Work by
G.K. was supported by U.S. Department of energy, Office of Science,
Basic Energy Sciences as a part of the Computational Materials Science
Program. K.M.S., A.W., and J.v.D. acknowledge support
from the excellence initiative NIM; A. W. was also supported by
WE4819/1-1 and WE4819/2-1.

\textbf{Author contributions}: X.D.,  K.M.S. and G.K. proposed this project;
X.D. performed the DFT+DMFT calculations and analyzed the results together
with G.K. and K.H.; K.H. developed the DFT+DMFT  code used and 
and assisted the computation setup; K.M.S. performed the DMFT+NRG calculations;  A.W. developed the NRG code and assisted K.M.S. in the initial stages of the DMFT+NRG computation. X.D.  and K.M.S. drafted the manuscript with
the help of G.K., K.H., A.W and J.v.D. 

\JVD{\textbf{Data availability:} The authors declare that the
data supporting the findings of this study are available within the paper [and its supplementary information files].}

\end{document}